# IT SECURITY PLAN FOR FLIGHT SIMULATION PROGRAM

## DAVID HOOD[1] AND SYED (SHAWON) RAHMAN, PH.D.[2]

[1]Information Assurance and Security, Capella University, Minneapolis, MN, USA
`dhood6@capellauniversity.edu`
[2]Assistant Professor, University of Hawaii-Hilo, HI, USA and
Adjunct Faculty, Capella University, Minneapolis, MN, USA
`SRahman@Hawaii.edu`

## Abstract

*Information security is one of the most important aspects of technology, we cannot protect the best interests of our organizations' assets (be that personnel, data, or other resources), without ensuring that these assetsare protected to the best of their ability. Within the Defense Department, this is vital to the security of not just those assets but also the national security of the United States. Compromise insecurity could lead severe consequences. However, technology changes so rapidly that change has to be made to reflect these changes with security in mind. This article outlines a growing technological change (virtualization and cloud computing), and how to properly address IT security concerns within an operating environment. By leveraging a series of encrypted physical and virtual systems, andnetwork isolation measures, this paper delivered a secured high performance computing environment that efficiently utilized computing resources, reduced overall computer processing costs, and ensures confidentiality, integrity, and availability of systems within the operating environment[1].*

## 1. Introduction

Developing a security plan for an organization as structurally secured and complicated as the Department of Defense (DoD) and the United States Air Force (USAF), is not easy for any organization. To make matters more difficult the DoD and the USAF have elected AISC to develop a flight simulation program and system to address pilot concern and alleviate safety concerns regarding pilot safety during simulation missions. AISC has addressed this requirement by developing a software based simulation system similar to those used in the space program.

---

[1]**Disclaimer**: The contracting vehicle mentioned in this chapter, Aviation Information Services Corp. (AISC), is a fictional organization and any mentioning of guidelines, processes, procedures, or policies are purely fictional and do not reflect upon current or past business practices utilized by the Department of Defense or the US Air Force.

This will eliminate any concerns regarding pilot safety and also eliminate safety and damage risks to both planes and pilots. With regards to information security, AISC had to consider classification levels and network isolation measures to ensure that the information and data housed within this environment was secured to the highest degree. Those challenges and implementation methods are described throughout this entire security plan.

This paper will outline the requirements for a secured computing environment leveraging methods of encryption and data protection while ensuring high availability of system resources. It will also outline vulnerabilities and attack surfaces that were identified and remediated to ensure that service level agreements were met. It will discuss the core service operating environment design and outline issues encountered during the design phase and how those obstacles were addressed. Once those issues are outlined, addressed, and remediated, change management is discussed and the overall project lifecycle is explained utilizing the software/system development life cycle as a baseline model. Lastly, this security plan will outline database systems utilized within the operating environment and how database and query security (to include scripting and coding vulnerabilities) are addressed and successfully overcome to ensure the system was delivered on schedule.

## 2. Background Study: Aviation Information Services Corp.

Aviation Information Services Corp, (AISC) is a consulting and contracting company that has been awarded a contract with the Air Force to develop a new aviation product that will be used in flight simulators to help train and develop pilots on new combat and maneuvering methods. This system will include secured methods of ensuring that this proprietary system (to include its information) and all systems that connect and/or rely on it for service availability are also adequately protected.

AISC is an organization with a centralized headquarters building with approximately 1500 employees. The headquarters building has about 250 employees located centrally where the remaining employees are contracted out to various projects and programs throughout the Department of Defense. ASIC is structured in two various methods; the core company has an executive group consisting of a Chief Information Officer (CIO), Chief Financial Officer (CFO), Chief Operating Officer (COO), and Chief Executive Officer (CEO). There are work sections within the organization as well that are managed by senior management personnel spread throughout the organization and their supporting contracts.

Within the COO tier (operations), there is a contracts program office that manages the various contract vehicles awarded to AISC. Within that structure is a sub-organization managed by a program manager and those individuals report up to the COO of AISC. This is because the contracts themselves are IT based and thus there is a more robust structure within the IT Services section of the contract vehicle. This is because the program itself is viewed as a separate organization. It still falls under AISC leadership and the executives are definite stakeholders in the program's success. However, those same stakeholders are not part of the internal contract decision making processes. Those only occur with either the COO or the contracts program management. This avoids convoluting your contract vehicle with too many management personnel interfacing the customer with various objectives. Keeping the  solution simple and limiting the number of contract executives helps ensure communications that pertain to the contract are kept inside a small circle of representatives within AISC and later briefed up the chain of command if need be.

# 3. Data Encryption Practices

## 3.1 Cryptographic System

The Department of Defense (DoD) has utilized a variety of secured access means with a variety of systems. Regardless of where the organization resides (agency, military, or otherwise), public key infrastructure (PKI) seems to be a common standard of secured systems deployment using methods of cryptography. The key difference between PKI systems in one organization compared to another is the cryptographic algorithm hash that is used to secure certificates. A cryptographic hash algorithm is a hash based algorithm that is designed to achieve certain security properties. The Federal Information Processing Standard 180-3 outlines the Secure Hash Standard and it specifies the five cryptographic hash algorithms; SHA-1, SHA-224, SHA-256, SHA-384, and SHA-512 for federal use. This standard has continued to be widely adopted by the information technology (IT) industry (CRSC, 2008.).

Within the DoD most environments use SHA-1 as a method of authentication against web services, domain services, and e-mail access and connectivity. Though not as secure as the other algorithms, there are DoD services that use this algorithm hash that are dependent on it and current tests are still ongoing as to whether these services are compatible with the more secure hashes. One example is the common access card which is used to store certificates used to login to DoD workstations and access secures web services and e-mail within the organization. These certificates are stored on the card themselves and they are used like PKI in e-mail programs and are used to secure and digitally sign e-mail messages to ensure non-repudiation of message transmission (CAC, 2011).

Since AISC's will be deploying this system within a DoD installation, it will adhere to all DoD standards regarding information security and technology. This solution will leverage DoD PKI configurations and utilize CAC cards to authenticate against the system. However, the system will also incorporate role based access control and limit access to the system with those that also have a dedicated PKI certificate which will be managed by a separately delivered PKI server that will house certificates of employees, contractors, and other personnel that will require access to the flight simulation system and its software. The method of access will start with the user using their CAC card to access the system using their SHA-1 hashed certificate housed on the card. That information will pass against a security policy that will be defined for the simulation servers and require an additional PKI housed on the PKI server. Those PKI certificates will be issued manually by a certification authority based on the users' requirement to access the servers. If they have both, they will be granted access to the system. From there, they will need to be members of the appropriate security groups to propagate access permissions based on the users' role and responsibility.

## 3.2 Encryption Mechanisms

The DoD has utilized a variety of secured access means with a variety of systems. Regardless of where the organization resides (agency, military, or otherwise), PKI seems to be a common standard of secured systems deployment using methods of cryptography. The key difference between PKI systems in one organization compared to another is the cryptographic algorithm hash that is used to secure certificates. A cryptographic hash algorithm is a hash based algorithm that is designed to achieve certain security properties. The Federal Information Processing Standard 180-3 outlines the Secure Hash Standard and it specifies the five cryptographic hash algorithms; SHA-1, SHA-224, SHA-256, SHA-384, and SHA-512 for federal use. This standard has continued to be widely adopted by the information technology (IT) industry (CRSC, 2008.).

Another encryption mechanism that will be utilized is SSL (secure socket layer) in conjunction with SFTP (secure file transfer protocol). These protocols provide a secure tunnel for data to traverse internally while also allowing file transfer to occur securely as well. S SL uses a program layer located between the Internet's Hypertext Transfer Protocol (HTTP) and Transport Control Protocol (TCP) layers. SSL is included as part of both the Microsoft and Netscape browsers and most Web server products. Developed by Netscape, SSL also gained the support of Microsoft and other Internet client/server developers as well and became the de facto standard until evolving into Transport Layer Security. The "sockets" part of the term refers to the sockets method of passing data back and forth between a client and a server program in a network or between program layers in the same computer. SSL uses the public-and-private key encryption system from RSA, which also includes the use of a digital certificate (SearchSecurity.com, 2011). SFTP, or secure FTP, is a program that uses SSH to transfer files. Unlike standard FTP, it encrypts both commands and data, preventing passwords and sensitive information from being transmitted in the clear over the network. It is functionally similar to FTP, but because it uses a different protocol, you can't use a standard FTP client to talk to an SFTP server, nor can you connect to an FTP server with a client that supports only SFTP (Indiana University, 2011).

AISC will leverage communication protocols like SSH, SSL, TLS, and SFTP are going to be utilized to securely encrypt data traffic and files that are utilized within the infrastructure. Though SSL are commonly utilized within web facing services, this will still securely protect data internally despite the fact that there is not an external burb configured on the internal firewall. This secured design is in place in the event that the DoD wishes to expand the flight simulation project to allow data to enter into the environment from externally facing resources within the DoD. Additionally, this solution will leverage DoD PKI configurations and utilize CAC cards to authenticate against the system. However, the system will also incorporate role based access control and limit access to the system with those that also have a dedicated PKI certificate which will be managed by a separately delivered PKI server that will house certificates of employees, contractors, and other personnel that will require access to the flight simulation system and its software. The method of access will start with the user using their CAC card to access the system using their SHA-1hashed certificate housed on the card. That information will pass against a security policy that will be defined for the simulation servers and require an additional PKI housed on the PKI server. Those PKI certificates will be issued manually by a certification authority based on the users' requirement to access the servers. If they have both, they will be granted access to the system. From there, they will need to be members of the appropriate security groups to propagate access permissions based on the users' role and responsibility.

One mechanism that will not be utilized within the flight simulation system is VPN. VPNs leverage the public network (i.e. the Internet) to access and securely transmit private data. This is done through a process called data tunneling which takes data traversing the internet and encrypts it so that it can travel safely and securely to the corporate infrastructure. This environment leverages secured systems and it is against DISA policy to allow VPN's to exist within a secured environment. Therefore, because VPN's are not going to be utilized within the system, other encryption protocols like point to point tunneling protocol (PPTP), layer two tunneling protocol (L2TP), and IP security (IPSec) will not be utilized. The key differences between PPTP, L2TP, and IPSec are the strength of the encryption hash that is utilized to secure the data transmission. Of the three, IPSec is the strongest using a 168 bit Triple-DES encryption (Raval& Fichadia, 2007).

## 3.3 Trusts in the Security Realm

As technological advances were made, businesses around the world had to adapt many of their practices around technological change. One area that changed drastically is how trusts are established in organizations. For example, with telecommuting the need to ensure that remote users are allowed authorized access to the infrastructure required additional security measures to ensure that there was authorized access by a trusted organizational resource. With the internet, this is more common since many web based businesses don't actually interface with their customers (or the customers with the business) (Raval & Fichadia 2007).

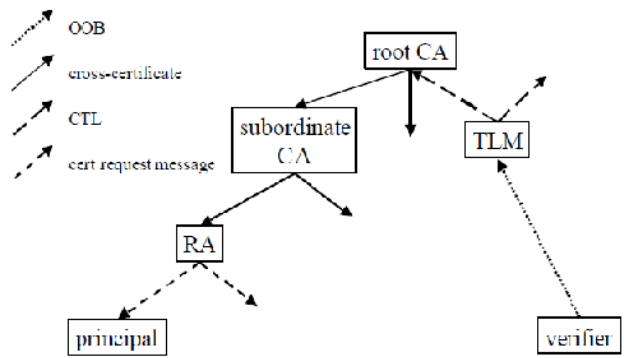

**Figure 1 – Hierarchical Trust Model (Moses, 2003)**

Therefore, these organizations take pride in making sure that their infrastructures are more secured to the organizations target level of trust. This trust level is set by the organization and is achieved by how their organization is secured. The higher the trust level, the more secure the environment. This depends on the business model and organizational requirements (for example PCI compliant businesses would have a higher trust level than a competing organization that was not PCI compliant) (Raval & Fichadia 2007).

## 3.4 Trusts in a PKI Environment

Within the PKI environment there are two major form of trust models; hierarchy and bridge. The hierarchical trust model is one in which ever key can be the subject of no more than one certificate or certificate request message. Within the hierarchical trust model there are four different trust mechanisms. It uses an out-of-band mechanism (OOB) between the verifier and the trust list manager, a certificate trust list (CTL) between the trust list manager (TLM) and the root certificate authority (CA), it also uses cross-certificates between the root CA and the subordinate CA and between the subordinate CA and the registration authority (RA) and a certificate message between the registration authority and the principal (Moses, 2003). Figure 1 (above) illustrates the trust mechanisms in a hierarchical trust model.

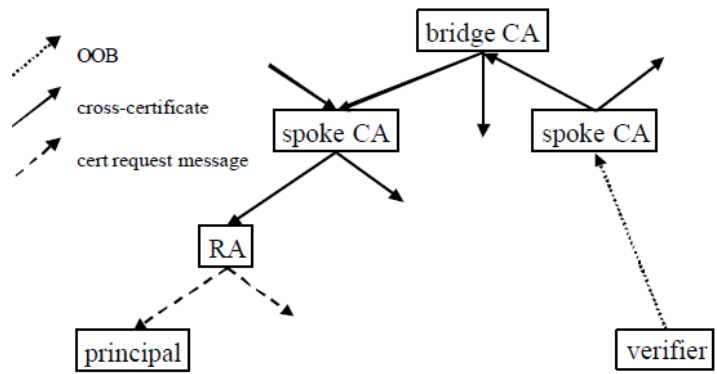

**Figure 2: Bride Trust Model (Moses, 2003)**

In the bridge trust model the subordinate CA can certify more than one root. In a hierarchical trust model, this isn't feasible because the subordinate CA would have more than one superior, which is inconsistent with the definition of a hierarchy. The bridge trust model has some similarities to the hierarchical trust model. However, the key differences is that the root CA is now the bridge CA and the trust list manager (TLM) is changed to the spoke CA, the same applies for the subordinate CA as well. The bridge trust model does not use the certificate trust list (CTL) mechanism but instead uses cross-certification. The largest difference between hierarchical and bridge trust models would be that the spoke CAs can be certified by more than one bridge, whereas in a hierarchy, subordinate CAs can only be certified by a single root CA (Moses, 2003). Below is a diagram illustrating the bridge trust model.

# 4. Application of Cryptographic Methods and Securing Systems

## 4.1 Disguising and Protecting Data

Regardless of which trust model you decide to implement into your environment, the concept of data protection and encryption remain the same within PKI. This is due largely to the fact that the components that make up PKI are the same, regardless of whether you utilize a bridged or hierarchical trust model (which really just differentiates whether you have a Root CA or a series of bridge CAs). The major components that make up PKI are the public key cryptography (PKC) and the certification infrastructure. The certification infrastructure provides for the creation, storage, and communication of digital certificates; which is managed by the certification authority. These certificates use PKC in the use of digital signatures (these signatures ensure the unique identifier of the sender or recipient of encrypted data) (Raval & Fichadia 2007).

With trust levels, it's very important in any PKI to understand and trust who is sending and receiving data. However, with PKI there is a major issue because public keys are published and can be utilized by anyone if they get a copy of it. For example, if user A sends a secured message to user B and that message is intercepted by a hacker and that hacker has user A's public key, that hacker can decrypt user A's message and modify it, sign it and send it back to user B. However, because user B doesn't have the hacker's public key, the message won't decrypt once user B receives it. The only way (and here is where the issue resides), that the message could get decrypted is if the hacker sends its public key to user B but disguised as user A. Then user B could save that key and now decrypt intercepted and modified messages from the hacker as opposed to user A. The concept of confidentiality and nonrepudiation are now flawed with the interception of a few messages because public keys are published resources (Raval & Fichadia 2007).

Trust models like the ones documented and discussed in Figure 1 and 2, can help improve the methods involved in disguising and protecting information and data using certification infrastructure. This is because the certification infrastructure consists of a chain of CA's; which are responsible for certifying the identity of an entity based on the entity's public-private key pair. Thus, any key obtained through this source can be technically considered as a genuine public key of the entity from which you have received a signed message (Raval & Fichadia 2007).Thinking back to our example, if user B had obtained user A's public key from a CA, the likelihood of user B saving a bogus public key would be reduced because the hacker is not a CA and therefore, cannot produce the authentic public key of user A.

PKI makes it possible for entities to sign messages using private keys (i.e. digital signatures). Once sent, the recipient uses the authentic public key of the sender (retrieved through a certificate authority) to decrypt the message using certification infrastructure. This digitally signed message is proof that ensures nonrepudiation.

## 4.2 System Hardening

The DoD and AISC both take systems hardening very seriously. Hardening measures help reduce the security footprint of a system and can also reduce the probability of attack without installing or configuring third party software. Hardening is the process of ensuring all security updates, patches, and configurations on a given system are applied to ensure the highest levels of security; thus reducing the systems security footprint on the infrastructure. These measures could be ensuring the latest patch levels are applied and locking down unnecessary services and protocols that aren't required by the system in order for it to function properly. AISC will use the Defense Information Systems Agency (DISA) Gold Disk standard in conjunction with vulnerability assessment scanning to ensure proper hardening of system resources slated for delivery under the flight simulation project.

DISA Gold is a software application that is constantly updated by a DoD managed patch repository system that runs on a system and checks for patch levels and unsecured protocols, services, processes, and configurations. The application includes various hardening methods (less secured unclassified configurations to very secure classified system settings), and they can run on any operating system approved for use on DoD networks (DISA, 2011).

Vulnerability assessment scans are accomplished after initial hardening is done using DISA Gold. Once a report of findings is reviewed by Information Security engineers and appropriate actions are taken against the findings, the vulnerability assessment scan uses a product called Retina. This product is a vulnerability assessment solution that integrates assessment, mitigation, protection, and reporting into a complete offering with optional add-on modules for configuration auditing, regulatory reporting, and integrated patch management. Retina enables you to centrally manage organization-wide IT security from a single, web-based console (eEye Digital Security, 2011).

AISC plans on mirroring the hardening techniques utilized by the DoD and the USAF with regards to hardening on all systems and products delivered under the flight simulation project. Given the nature and hybrid infrastructure of the USAF and the DoD, the hardening recommendations and reviews will have to be conducted manually by systems integration engineers and information security engineers to either mitigate, waive, or justify (assuming the risks) of findings conducted during both DISA Gold scans and vulnerability assessment scans. Additionally, scans will be conducted during security and accreditation (C&A) renewal as necessary. This is important as there are many vulnerabilities that reside within many operating system environments. An example graph is shown below.

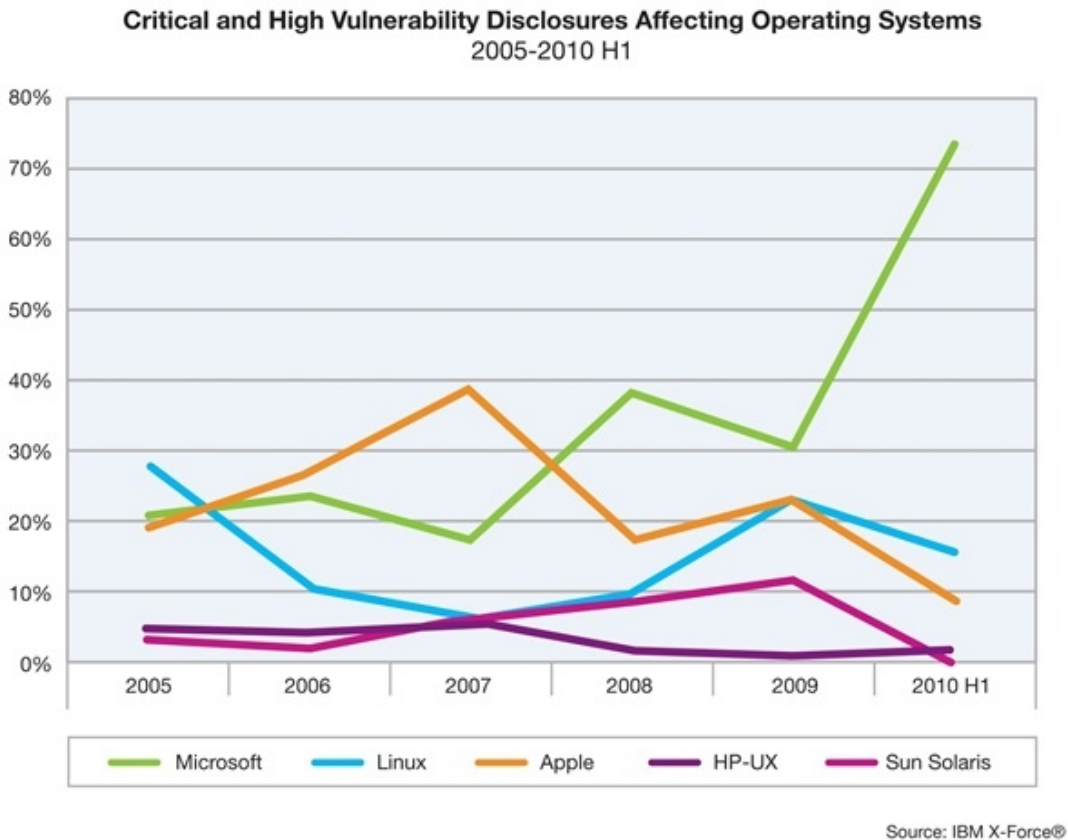

**Figure 3 –Operating System Vulnerabilities (IBM, 2010)**

This chart outlines the percentage of critical or high number of vulnerabilities that have existed within Microsoft, Linux, Apple, HP-Unix, and Sun (Oracle) Solaris (Unix) operating system environments over the past 5+ years. In today's infrastructure Microsoft operating system environments make up for approximately 75% of high and/or critical vulnerabilities (IBM, 2010). This is important to note as a large majority of client workstations and server systems in businesses today are Windows based. The importance of system hardening (be that a remediation server, scanning policy, or automated patch management system) is critical to ensure that your systems reduce their security footprint as much as necessarily possible.

## 4.3 Confidentiality, Integrity, and Availability

AISC as company has a centralized security architecture with a consolidated data center hosting all the corporate IT services for the organization at the headquarters building. There are centralized security measures like access rosters, alarm systems, biometric access to the data center and methods of encrypted data access and authentication. These solutions are all geared around the CIA Triad methodology of security where the framework consists of three major components (confidentiality, integrity, and availability) (Stallings & Brown, 2008).

From a confidentiality standpoint, AISC works very closely with the DoD to ensure that security classifications and need to know is enforced on programs and special projects. Within the IT realm, this means data privacy and protection. Within both the DoD and USAF, data integrity and protection is of the utmost importance with the growing threats of cyber-attacks. It is this exact reason that AISC has developed the flight simulation program in an access controlled hanger where systems are segregated and consolidated off of the enterprise network and thus ensuring protection from unknown and outside resources.

Integrity with the flight simulation project is integral because this means that the system resources within the project are performing their desired and engineered functions as designed (without problems and interruption). This goes in line with governance of policy and procedures on systems integration and information security measures within a given design. Given the nature of the project policy and directional changes don't happen within AISC but within the USAF, DoD, or DISA (for larger governance direction). AISC has recommended additional integrity measures within the data center that houses the flight simulation project servers (to include equipment and access control measures).Additionally, theequipment and accesses which includes access control rosters, restricted access permissions based on roll based access control (RBAC), and systems access control lists are also listed as additional security measures.

As it pertains to availability, AISC will leverage redundant power and generators within the military installation to ensure that power is always available to the simulation equipment, servers and other hardware required. Within the system design itself; all server hardware will be configured with redundant components (memory, fans, processors, power, etc...) to ensure system availability. Additionally, from the transport (networking) will be redundant from the server systems to the switches. The switches, routers, and other transport devices will also have redundant systems and connectivity to other systems to ensure high availability of all hardware.

However, in the event of a server failure, AISC has also designed the servers to be highly available leveraging a virtual infrastructure which will consist of approximately four physical servers and 25 virtual machines. These virtual machines will distribute evenly across three of the four physical servers where one server will be a dedicated failover host in the event of hardware component failures. Given that AISC prides itself as a technology delivering organization, effective and efficient systems design and high available solutions are vital to their success. Additionally, the higher availability that a system or service has, the more satisfied the customers are going to be.

# 5. Cipher Attacks and Risks to Information Systems

## 5.1 Protection from Malicious Ciphers

Both the DoD and AISC understand the importance of information security. Given the nature of what the DoD does and the sensitivity of many of the programs that fall within the scope of the DoD, ensuring that all pieces of technology are secured to the highest level while ensuring high availability of services and connectivity. Ensuring that your environment is protected from malicious code or other attacks is vital to the success of any organization. To the DoD and the military, it can save lives.

AISC takes the DoD's stance on security very seriously and in order for them to effectively build a foundation of protection from malicious ciphers; they need to establish excellent transport security. Transport security protects communication from being exposed to un-trusted third-parties that try to attack or compromise the system in order to steal information; not protecting this communication only exposes your organization to potential attacks (Auger, 2010). Web services hosted within the DoD and AISC use SSL/TLS to ensure web services and the communication that traverses between users and the websites are secured. Additionally, these web sites use much stronger cryptographic ciphers used to protect information (OWASP, 2011).

This will be the same process for securing the simulation systems and services being developed by AISC. The system will provide support for only strong protocols and certificates against services and web services will only support domain names that end in .mil or .gov. Because this application interfaces with the defense network, there is no reason for .com access to these systems since they are proprietary to the installation and the organization that utilizes them.

## 5.2 Cipher Attacks and Defense

Both the DoD and AISC understand the importance of information security. Given the nature of what the DoD does and the sensitivity of many of the programs that fall within the scope of the DoD, ensuring that all pieces of technology are secured to the highest level while ensuring high availability of services and connectivity. Ensuring that your environment is protected from malicious code or other attacks is vital to the success of any organization. To the DoD and the military, it can save lives.

AISC takes the DoD's stance on security very seriously and in order for them to effectively build a foundation of protection from malicious ciphers; they need to establish excellent transport security. Transport security protects communication from being exposed to un-trusted third-parties that try to attack or compromise the system in order to steal information; not protecting this communication only exposes your organization to potential attacks (Auger, R., 2010). Internal web services hosted within the DoD and AISC use SSL/TLS to ensure web services and the communication that traverses between users and the websites are secured. Additionally, these web sites use much stronger cryptographic ciphers used to protect information (OWASP, 2011). Despite the above mentioned methods of protection, we still have to monitor and report these systems in real time in the event of an attack from within the infrastructure. This could be in the form of malicious code being introduced within the environment or data corruption occurring within applications inside the infrastructure. The flight simulation project is uniquely different from other products because there is no web facing connectivity within any system. This is because there is no external burb configured within the firewall. This significantly reduces the many types of cipher attacks that could occur from outside the organization. However, if the DoD configures the firewall to communicate against the internet they will have to review and defend against cipher attacks like brute force or linear cryptanalysis.

Brute force attacks are attacks that systematically attempt every possible key. More often than not it is used in a known plaintext or cipher-text only attack. These attacks are dangerous because they are always successful; given they have a finite key length and sufficient time. Methods of defense against brute force attacks would be the use of advanced encryption standard (AES). This is significantly more secure than DES and takes significantly longer to compromise. Couple this encryption technology with a good protocol sniffing utility and intrusion detection system (both are incorporated into the monitoring and security suite systems), you significantly reduce the potential of a successful attack (Conrad, 2007).

Linear cryptanalysis is a known plaintext attack that requires access to large amounts of plaintext and cipher text pairs encrypted with an unknown key. It focuses on an analysis against a single round of decryption on a significant amount of cipher-text. The individual behind the attack decrypts each cipher-text using all possible sub keys for each round of encryption and studies the resulting intermediate cipher-text to seek the least random result (Conrad, 2007).  Common methods of protection against these types of attacks are strong monitoring solutions and strict monitoring of intrusion detection systems.

There are numerous issues that can arise with both secured and unsecured application systems within the flight simulation project. Leveraging encryption methodologies and technologies like SSL, SSH, SFTP, TLS, and PKI, AISC will guarantee that information and traversing the wire from system to system is secured and only accessed by authorized resources. Additionally, utilization of RBAC and two-factor authentication systems like Active Directory with digital certificates via a CAC card and PKI certificates will guarantee that information is not only access using something that someone knows but something that they have and/or are (to include the use of biometrics to access secured systems).

# 6. Systems Architecture and Application Design Issues

## 6.1 Security Systems Architecture

AISC as company has a centralized security architecture with a consolidated data center hosting all the corporate IT services for the organization at the headquarters building. There are centralized security measures like access rosters, alarm systems, biometric access to the data center and methods of encrypted data access and authentication. These solutions are all geared around the CIA Triad methodology of security where the framework consists of three major components (confidentiality, integrity, and availability) (Stallings & Brown, 2008).

However, as a contract on various defense agency programs, their security architecture is more hybrid. Hybrid technologies are a mesh of both centralized (consolidated security architecture generally in one single site), and decentralized (security architecture spread all over the place (country, world, etc…) (FreeOnlineResearchPapers.com, 2011). Since the contracts awarded to AISC are to the DoD (currently to the USAF), the security architecture would be that of the organization AISC is working as a contract for. The USAF and the DoD have a large centralized architecture with primary sites at various network operations and security centers (NOSCs). However, each individual installation or base has its own architecture that communicates up to the NOSCs and in some cases to other agencies (depending on what missions are being supported at that installation).

With regards to their recent contract where they will be developing a new aviation program that will be used in flight simulators to help train and develop pilots on new combat and maneuvering methods, the architecture is hybrid as well. This is largely due to the fact that the USAF itself is a very broad and robust network with both centralized and decentralized architectures (i.e. deployed locations like Afghanistan and Iraq).

## 6.2 Secure and Unsecure Application Issues

The biggest differences between AISC's flight simulation project design and other system designs are that they are developing a secure system for the department of defense (DoD); and that this environment will not be publically facing (meaning there is no access to the Internet from within the system). Because of this even unsecured application issues are minimal at best because most security concerns around unsecured applications are stemmed from attack and compromise from outside attacks (i.e. hackers). However, the biggest issue is ensuring proper access control measures are executed properly to guarantee that least privilege is enforced at all levels of the system. This will provide a solid security framework for accesses to resources within the project.

To recap, Figure 4 is a diagram used to reference some of the application systems that are required to access, control, and operate the flight simulation project. As you can see in Figure 4, there is no link to the outside (Internet). However, there is a design in place to establish a link if the DoD later requires access to this system from the outside (configure the firewall external burb to access the externally facing routing device). Within the application architecture of this system the primary applications utilized are active directory, the database application, security suite application, development tools, PKI certificate application, monitoring application and tools, and the flight simulation application that will process all the intelligent code and graphics necessary to

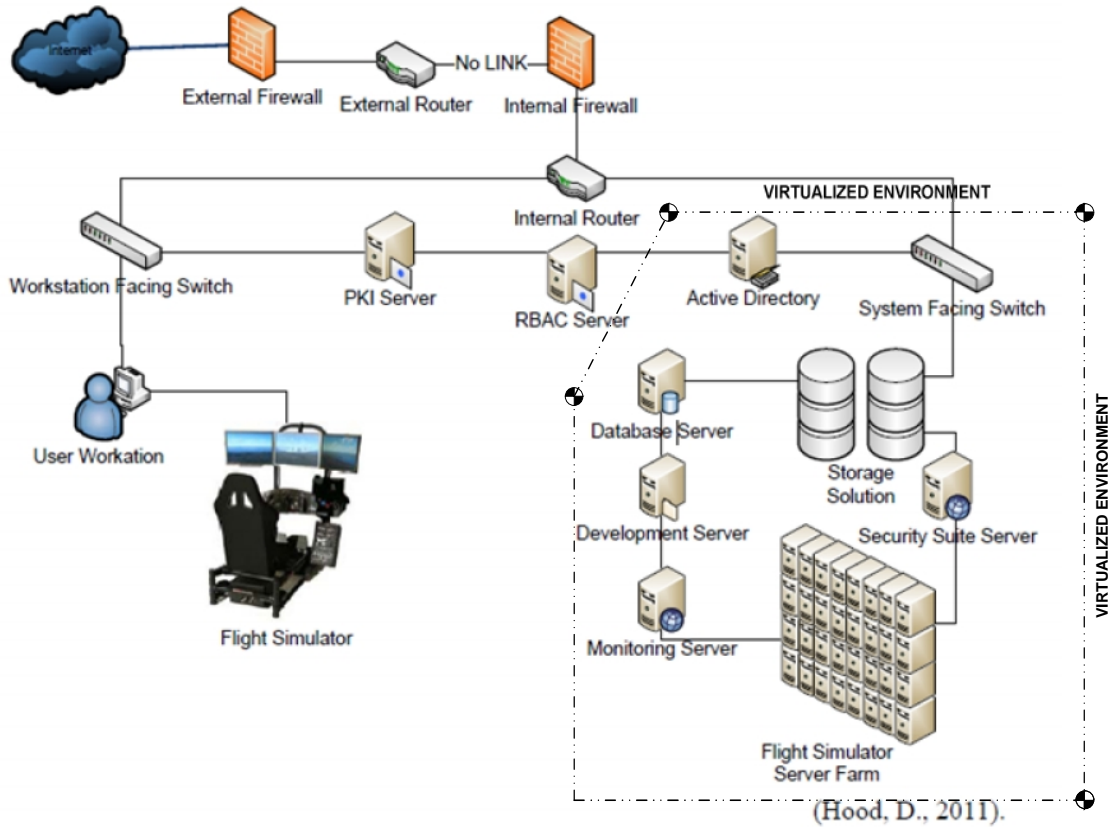

(Hood, D., 2011).

display an environment on the flight simulator itself. These environments all process various levels of secured information and each individual system is using a method of security to access the system (via discretionary access control lists, role based accesses, and two-factored authentications). Note, in the below diagram that storage is accessed via virtual systems but the storage SAN solution itself is not virtualized.

The largest issue currently facing AISC and the system design teams is the exercising of least privileges on various role based access controls. Due to the security complexity of the system set forth via requirements by the DoD, AISC has a difficult task on developing a secure system while ensuring proper access controls are established for the design team. This causes confusion during configuration of various applications because certain applications require higher privileges on one system over another. A great example is any application that is writing to the database server cluster. Even though the systems engineer has administrator privileges in Active Directory and the application being designed uses active directory credentials to authenticate against it, it also writes data to the database. Given that the database is Oracle and doesn't immediately authenticate or replicate against Active Directory there has to be permissions granted to the user at the database system. For engineering and design purposes this isn't increasingly difficult to accomplish or justify. However, it must be audited and controlled so at the completion of the project, all unnecessary permissions, privileges and accesses granted can be appropriately removed.

Another issue is role based assignment on the server operating system. The initial design used a bare instance of Windows operating system but the DoD required the use of their approved and accredited image of the OS. Therefore, AISC was tasked with imaging (a process of building systems off a particular pre-approved design) all the servers. However, at the time of deployment it was noticed by engineering staff that servers imaged included web services on these systems. Though web traffic is not enabled at the internal firewall, if later addressed and established, could cause significant security concerns and potential web exploitation from the outside. Therefore, engineers raised these concerns and all unnecessary web services were disabled on the servers.

## 6.3 Secured Design and the CIA Triad

AISC as company has a centralized security architecture with a consolidated data center hosting all the corporate IT services for the organization at the headquarters building. There are centralized security measures like access rosters, alarm systems, biometric access to the data center and methods of encrypted data access and authentication. These solutions are all geared around the CIA Triad methodology of security where the framework consists of three major components (confidentiality, integrity, and availability) (Stallings & Brown, 2008). From an application standpoint, this also includes the use of encryption protocols like SSL, TLS, and SFTP to guarantee that access (both physical and virtual) are secured from the moment of entry to exit.

### 6.3.1 Confidentiality

From a confidentiality standpoint, AISC has to work very closely with the DoD to ensure that security classifications and need to know is enforced on programs and special projects. Within the IT realm, this means data privacy and protection using the above mentioned encryption protocols in addition to additional access control measures like ACLs and RBAC measures. Most of the application systems within the flight simulation project are Windows based and Within Windows there are two kinds of access control lists; discretionary access control lists (DACL) and System ACL (commonly referred as just an ACL). When most people use the term ACL they are more than likely referring to a DACL; this is where one is granted or denied access to protect resources in Windows such as files, memory, and other resources. Examples of these types of access lists would include the Power Users or Administrators Security group. These groups provide special access to resources like files and services. Additionally, you can perform more advanced functions like download and install certain programs and applications (Stallings & Brown, 2008). Within both the DoD and USAF, data integrity and protection is of the utmost importance with the growing threats of cyber-attacks. It is this exact reason that AISC has developed the flight simulation program in an access controlled hanger where systems are segregated and consolidated off of the enterprise network and thus ensuring protection from unknown and outside resources.

### 6.3.2 Integrity

Integrity with the flight simulation project is integral because this means that the system resources within the project are performing their desired and engineered functions as designed (without problems and interruption). This goes in line with governance of policy and procedures on systems integration and information security measures within a given design. Given the nature of the project policy and directional changes don't happen within AISC but within the USAF, DoD, or DISA (for larger governance direction). These directives not only pertain to policy and procedures but also to configuration management of the design and build of the systems and their applications; thus the inclusion and enforcement of encryption methods and protocols. AISC has recommended additional integrity measures within the data center that houses the flight simulation project servers and equipment which includes access control rosters, restricted access permissions based on RBAC, and systems access control lists (Stallings & Brown, 2008).

### 6.3.3 Availability

As it pertains to availability, AISC will leverage redundant power and generators within the military installation to ensure that power is always available to the simulation equipment, servers and other hardware required. Within the system design itself; all server hardware will be configured with redundant components (memory, fans, processors, power, etc...) to ensure system availability. Applications hosted on the server environments will be redundant as well. As shown in Figure 4, each system displayed is a representation of a clustered environment. This means that systems are configured using multiple servers and clustering solutions to provide an active-active or active-passive environment. This guarantees high availability of all services at the application level. Additionally, from the transport (networking) will be redundant from the server systems to the switches. The switches, routers, and other transport devices will also have redundant systems and connectivity to other systems to ensure high availability of all hardware.

However, in the event of a server failure, AISC has also designed the servers to be highly available leveraging a virtual infrastructure which will consist of twenty physical servers and one hundred virtual machines, though not immediately identified in Figure 4 (Data Flow Diagram), the application servers are all hosted within the server farm as a virtual guest operating system. The physical equipment is deployed as a hypervisor (creates the virtual layer for the operating systems to utilize resources and physical components). These virtual machines will distribute evenly across the rest of the physical servers. Each environment (development, test, and production) will be hosted on a dedicated cluster of resources dedicated to only that environment. Therefore, we'll have an adequate testing plan and pre-deployment plan included in our design. Additionally, we'll have plenty of resources necessary to host the graphics engine necessary to run the simulation program and the individual simulations that each pilot will utilize.

### 6.3.4 Quality and Reliability

When taking software quality, reliability, and security into consideration, AISC cannot focus too much of their attention on one specific area over another. The fact of the matter is that all three possess a very high importance to not only AISC but the DoD as well. Considering the fact that the flight simulation project is a DoD funded effort and that AISC is the primary contract vehicle, there are standards and requirements for quality, reliability, and security that must be incorporated into all applicable phases of the software development life cycle.

As it stands currently, the primary development code used to develop many of the simulation applications and programs within the simulator are going to be developed using PERL (scripting and web development), SQL (for database queries), and C++ (used to develop the simulation environment). All of these languages are going to require serious time and effort to develop especially given the strict security requirement and heavy importance on quality and reliability.To understand the importance of software quality, reliability, and security as it pertains to the flight simulation project, one must understand the systems and security architecture of how a user would interact with the flight simulator and how the simulator would interact with information within the specific simulation being run on the system. Furthermore, that data that resides in this server farm must be secured and coded as effectively and efficiently as possible with no errors.

As you can see in Figure 4, there are multiple incorporations of security layers to access the data housed within the flight simulator server farm. Each server is a dedicated cluster of those systems to ensure high availability and redundancy. The applications and programs will be loaded via a user workstation and controls within the simulation will be handled by the custom designed flight simulator machine. The user session is loaded via a secure login against a variety of access control measures. As previously stated, authorized users have to be members of specific role based groups with appropriate permissions. Additionally, they will have to have proper credentials against a directory server and also possess a PKI certificate.

However, the user environment will not load without a securely and reliable application that houses all the intelligent data necessary for the flight simulator to function effectively. However, the code that runs these applications has to run securely as well. Therefore, the development team will incorporate obfuscated code into the application development. Obfuscated code is code whose logic is intentionally difficult to follow and/or whose syntax is intentionally unclear (Dream-In-Code, 2007). This is going to cover two separate grounds, it will make the application more secure and reduce risk of specific code tampering or theft and it will protect intellectual property of both AISC and the DoD.

## 6.4 Change Management

When AISC was deploying the systems into the production network, one area that was initially overlooked was addressed at one of the larger technical issues encountered. During deployment some code was introduced into the flight simulator software and caused some issues with the applications. This unfortunately was not tested nor validated prior to introduction to the production network. Therefore, a change management process had to be introduced as new systems, code, and applications were introduced into the flight simulator environment.

This environment would be responsible for ensuring that there is a system baseline and that this baseline would be kept up to date and free from certain changes to the environment without pre-approval or testing procedures followed. Within many of the work centers at the USAF and the DoD, change management personnel ensure that systems are patched accordingly and applications and programs are tested prior to deployment into a production environment. More importantly, they ensure that there are written or documented (electronically) processes and procedures that outline the change management process and the system baseline in question.

The risks associated with not having a change management process are significant and if (like the above example) you introduce non-tested software into an environment as custom as the flight simulation project, you could cause significant damage to the environment that would result in hundreds of thousands of dollars lost and hundreds of man hours wasted in an instant.

# 7. Software Development Life Cycle (SDLC) Evaluation

## 7.1 Securing the Organization and the Systems Development Life Cycle

When viewing the organizational security policy on how to ensure the most effective and efficient solution to securing data transmissions, you have to look at the requirements; the same applies when developing systems using the systems development life cycle. In large environments that have multiple security policies and large data center environments (meaning multiple node clusters per system or service delivery), the hierarchical trust model will lead to duplication and redundancy of subordinate CAs and principals' private keys. The bridge trust model allows a single spike CA and single principal private key to be reused in more than one bridge (Moses, 2003). However, within each bridge, there is only one copy of the CA and principal private key. Thus if your trust is broken between bridge CA's, you lose redundancy with your CA and principal private key.

Hierarchical trusts are more secure in large infrastructures and allow for higher probability of trusted transmission of data between the sender and receiver of secured messages. Additionally, due to the fact that their hierarchical trust is redundant you have a trusted CA issuing and verifying the authenticity of the messages being sent or received. Within the user community the recipient can rest easy knowing the public keys being generated or sent are being done so by a trusted root or subordinate CA and not through some unauthorized source. Also, in the event of keys being misused or use of invalid or expired keys, the CAs also has certificate revocation lists (CRL) that can revoke key certificates. This also adds an additional layer of security by allowing certificates to expire.

PKI is designed to address several security concerns of the organization. With the use of a redundant hierarchical trust model with PKI your organization ensures that the following objectives are met: (Raval & Fichadia 2007).

- Authenticating the communicator – Successful authentication helps achieve nonrepudiation. To avoid mis-authentication the sender's public key needs to be obtained by a trusted source (i.e. root or subordinate CA).

- Ensure that the message the sender transmitted came to the recipient without modification – this is accomplished using the user's private key to encrypt the message (or message digest). This method secures the message integrity objective; anyone who decrypts the message using the sender's public key will not be able to re-encrypt it. At the same time, this provides proof of the sender's identity.

Like all projects there is a requirements analysis which the DoD performs just like any other business. AISC reviewed these requirements and the systems delivered were agreed upon by both organizations before a design was laid out. AISC worked more diligently with the DoD and the USAF during each phase of the SDLC and unlike other organizations that like to press through each phase upon its completion, AISC, DoD, and the USAF would (if necessary) revert back to previous phases to readdress situations or concerns that may have arisen during various phase of the project. Doing so allow all areas within that phases to be addressed and worked (and re-worked) to ensure the successful completion of that phase and   the project as a whole.

## 7.2 Revision Control and the SDLC

Revision control is the management of changes to data resources. It's commonly utilized in the software development realm where a group of individuals can change the same files. Changes are reflected in the order they were made by a letter or code (identified as a revision number) (BetterExplained, 2009). This allows excellent forms of data file accountability and ensures that people are working on the right files and that the development teams working on the programming code for the project have the latest and most up to date information handy.

Revision control systems (sometimes called version control systems (VCS), are great at letting users track files over a period of time. The reason that these environments are so handy within software development is because during the development (coding), testing, and development of documentation for a project, these files are not generally stored in a file share where multiple updates can be made simultaneously. If this were the case, you'd lose accountability of your data because multiple users would think that their updates to a file are the most recent. This is where the introduction of a VCS is a good idea so that you have that high level of accountability within your software development teams and the files they are creating and/or updating.

The most commonly utilized (and one that has been around a while) VCS is a product called CVS. This product was introduced in 1986 and it is the "de facto standard" and is installed in numerous places. It's a simplified VCS and isn't as fully featured as other products but this also makes CVS easy to learn and simple to use. It ensures that files and revisions are kept up to date and though it's older, it's still widely utilized within the development community for backing up and sharing files (Stansberry, 2008). AISC is utilizing CVS during the development, testing, and documentation phases of the flight simulation project to ensure that there is accountability of files amongst the development team and technical writing staff. This ensures that teams aren't stepping on anyone's toes during the project and ensures smooth transitions of each phase of the project.

# 8. Database Management and Database Security

## 8.1 Database Security Considerations

AISC is leveraging role based access control (RBAC) and PKI architectures (i.e. certificates) to grant and authorizes access to the systems within the scope of the flight simulation project. With respect to the databases and the DBMS, AISC will also leverage encryption of the data, the stored files, and the backups of the databases. This is due largely to the fact that the information stored within these databases is classified and the utilization of these methods helps maintain data security and access control of classified information. Within the DBMS access will be determined by a similar RBAC measure where role is determined by users affiliated with a job or position (i.e. database engineers and administrators) and those users are added to a security group that is given exclusive access to the DBMS; no one else is granted access to the DBMS (Stallings & Brown, 2008). This is because the databases are complex and standard users (even those with access and clearances to the entire system) don't require access to the databases or the DBMS.

Another consideration that AISC has taken into account is separation of database services from other services. Generally speaking, a best practice is to separate any databases and DBMS environments from publically facing (web) system (Orloff, 2011). This avoids attacks on a system (web server for example) that could also be housing your databases or DBMS. In this chapter, there is no mentioning of a publically facing environment because this system is completely separated from the defense network and is only used to ensure that the flight simulation programs and systems function as designed. This in itself makes the database environment more secure than other environments because this eliminates the threat of an unauthorized hack or compromise from an outside resource. The only risk to the system is going to be from within the walls of the room where the flight simulator and the systems are centrally located.

Even if your environment is locked down from physical and network access the potential can exist for incompatibilities to exist within your environment. The potential of corruption can still happen within the programs, databases, and applications (Orloff, 2011). This is why that all systems (especially the databases and DBMS) are patched appropriately and that there is a documented and enforced patch management policy (to include testing).

## 8.2 Following the Governments Lead

Though AISC has been awarded this contract to design this complex system, they still have to adhere to the government's rules and regulations concerning information security and technology implementation. This isn't any different for the databases and DBMS responsible for managing the data that is going to be responsible for the continued successful operation of the flight simulation project. The largest parameters that exist within the information technology are the compliance regulations that all DoD systems must fallow; DISA STIG compliance.

DISA (Defense Information Systems Agency) STIG (Security Technical Implementation Guides) are regulations that are utilized to enforce how systems, networking devices, computers, and communications appliances like firewalls and intrusion detection systems are deployed into an environment. There are STIG's for these devices of all security classifications and environments within the DoD (say an intelligence agency) (DISA, 2011). Even though this environment is separated and not connected to the defense network, all systems within the scope of the flight simulation project must follow these guides prior to being implemented. This also falls in line when it comes time to process the certification and accreditation (C&A) package of the environment. Therefore, creating a secured environment for use isn't just a goal of AISC; it is also a requirement of the government.

Creating a secured environment within the government's requirement is not a simple task even though you have a checklist of findings that have to be marked off. As the security classification increases; so does AISC's workload. However, some findings can be waived or justified (just like risks can be assumed) a great example of this would be the incorporation of a firewall appliance. The flight simulation systems environments (to include the databases) are not going to be communicating with the rest of the defense network. Therefore, a firewall appliance wouldn't be necessary (DISA, 2011). However, AISC looked to the potential of growth and connectivity to the defense network to publically and securely allow others to view flight simulation data and logs without the necessity of having to go onsite physically to view them from within the walls of the data center. AISC still incorporated both a firewall appliance and an intrusion detection system into the design plans. Though not enabled during the final implementation phase, these devices will allow the government to maintain security compliance in the event that they decide to connect this environment up to the defense network.

## 8.3 Database Fundamentals and Secured Database Requirements

One of the largest concepts of database management is the interoperability with the operating system environment. This is important as it pertains to access control measures operated within the database environment and (more importantly the DBMS). There are some functional differences between operating systems and DBMS. Operating systems tend to deal with subjects attempting to access some object. DBMSs are employed for sharing data between users and to provide users with a means to relate different data objects. Also, DBMSs are generally dependent upon operating systems to provide resources such as inter-process communication and memory management. Therefore, trusted DBMS designs often must take into account how the operating systems deal with security (Thuraisingham, B., 2008). Due to the interaction with the operating system environment and the sharing of resources within that operating system (i.e. disk, RAM, and processor), the DBMS will be housed on a very powerful LINUX environment. LINUX requires fewer resources to operate independently than Microsoft Windows and is also more secure than most Windows environments.

Another fundamental operation of database management is the use of monitoring and reporting of activities conducted on the database(s). Certain programs can help assist in the automation and publication of alerting administrators and engineers in the event of a problem. However, many skilled database engineers and administrators will be able to determine issues that have arisen based on recent behavioral changes that have occurred within the database or the DBMS (FFIEC, 2011).

To alleviate database engineering resources, AISC has deployed a system responsible for monitoring application services, processes, and system resources 24 hours a day. Additionally, these systems will monitor and audit changes made within the environment to provide accountability of all privileged users that have access to all systems. This is especially useful in secured database environments because it can monitor and alert on internal and external modifications that may have been made to the database environment and quickly alert those in the event of critical issues.

Another fundamental outlined earlier was regarding encrypted communication. Within the DBMS you can establish encrypted communication between user's access to the database and the databases housed within the DBMS. However, the connectors on the DBMS themselves are not necessarily as secured as they could be. Integrity lock is a security architecture concept that utilizes a trusted filter that is maintained at the highest security level supported by the DBMS. Using this method, every element that is keyed into the database is associated with a security label and a cryptographic checksums. From there, the checksums are computed by that trusted filter on insertion and recomputed during retrieval (Thuraisingham, 2008).

In addition to ensuring that the database and DBMS are performing their desired function, AISC strives to ensure that they are performing those functions as efficiently as possible. This is why in systems development; you have to account for proper resource utilization, management, and most importantly performance. As stated in previous project documents, AISC ensures the highest level of redundant resources available within the DBMS to ensure efficient operations of the databases. This guarantees data availability and system efficiency at all times because of its robust and highly available design. This also translates into a government requirement and guarantees data availability and data integrity.

One last concept and fundamental that outlines most database designs is the avoidance of dynamic SQL. Dynamic SQL are statements that are constructed by including user inputs directly into an SQL query. This is generally done with what Oracle calls "bind variables" (SQL Server calls it "parameterized SQL") wherein instead of directly inserting user-supplied input into the SQL query, inputs are first assigned to parameters (variables0) and then the parameters are used in the SQL query (Raval & Fichadia 2007). Dynamic SQL can open the environment to additional vulnerabilities like SQL injection, where malicious codes can be entered into a query from an unauthorized user or resource via the web (in most cases) (Raval & Fichadia 2007).

The interesting concept is the type of scripting languages that result in attacks, SQL injection make up about 18% of surface attacks as outlined below. AISC has deployed methods of monitoring the query inputs via managed systems and only authorized users granted the appropriate role, group access, and permissions will be able to enter queries into the database. In addition to not utilizing dynamic SQL, the flight simulation project will not be web facing and will eliminate the risk of unauthorized access via the web. As noted in Figure 7,you have the top ten classes of attack type. Each bar graph represents an attack vulnerability represented in the color coded legend. The percentage is measured as a likelihood of a website having vulnerability by fore mentioned attack vulnerabilities. In 2010, the top attack method was cross-site scripting. At the end of the year moving into 2011, information leakage moved in line with cross-site scripting and is slowing working past as the top attack class (White Hat Security Inc, 2011).

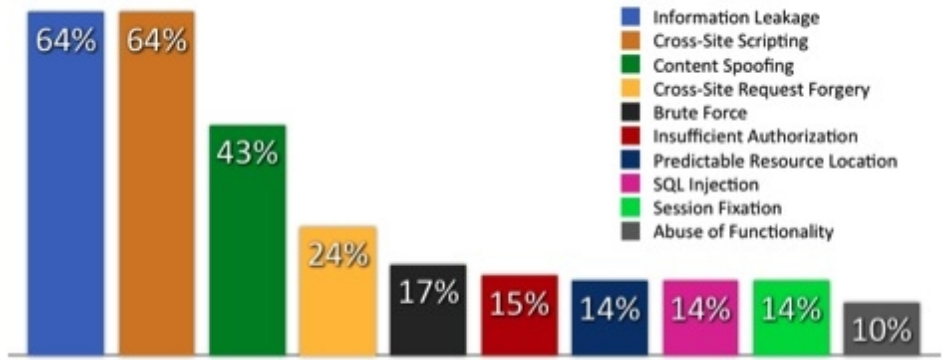

**Figure 7 – Top Ten Classes of Attack(WhiteHat Security, Inc., 2011)**

AISC knows the importance of information security as it pertains to databases and DBMS environments. Therefore, AISC will be utilizing the above considerations on the development of a secured database environment running Oracle 11g running within a Linux operating system environment. All communication traversing the wire to and from the DBMS will be encrypted and role based access control will be utilized in conjunction with PKI certificates to determine appropriate access control of the database environment.

Given the location of this environment and the security classification levels of the data that are going to be involved, AISC has to follow strict government guidelines and requirements when developing the project systems (to include the database environment). AISC will incorporate the project design and ensure that all systems, networking devices, operating systems, and applications are configured in accordance with DISA STIG regulations. Lastly, AISC will ensure that there is the highest level of data accountability by deploying a version control system (VCS). This ensures a smooth transition of each phase of the software development life cycle (SDLC).

# 9. Conclusion

AISC has developed a complex environment that interacts with numerous secured systems that leverage many government access control measures. This environment leverages complicated and lengthy programs that conduct many intelligent operations automatically. Developing a system in accordance to governance by the DoD and USAF, in addition to following a project software development lifecycle has been a taxing process for AISC engineers, and government personnel associated with the project.

This chapter has outlined the necessity and requirements for ensuring that levels of cryptographic systems and various levels of encryption are incorporated into various levels of the system environments. Additionally, the importance of cryptographic cipher utilization and risks associated with the deployment of unsecured system environments were also outlined. Given the environment that AISC was working in and their past experience in project work, developing an unsecured environment or system was not an option. Of course, no project was completed successfully without its list of issues. AISC encountered some issues with change management, process, and role based access assignment in the phases where systems development and application configuration were performed. This is where AISC especially learned the importance of a solid change management procedure and proper test-bed environment. Lastly, with the intelligence of this environment being as complex as it is, constant read/writes had to occur and complex queries had to take place within the database environment. Therefore, AISC had to work diligently on developing a secured database instance and database management system.

Despite the increased levels of security and encryption, the flight simulation project will still guarantee data and system confidentiality, integrity, and availability through the use of clustered operations and systems designed at the application, server, network, and storage system levels. Leveraging virtual technologies at the system levels will also add an additional layer of filtering and security to ensure confidential access to system resources. AISC is designing the flight simulation project to be Internet facing. Though there is no external burb configured at the internal firewall (meaning no internet facing connections), AISC has designed the entire project around the potential that the DoD will configure the burb thus making the system internet facing. This means that cipher attacks like brute force, meet in the middle, or linear cryptanalysis will be defended against using an array of intrusion detection methods, active constant monitoring of protocols within each system, and strengthened encryption keys.

## Authors

David L. Hood is a Senior Systems Engineer serving as a defense contractor within the defense intelligence community. He is also a student working on his Masters of Science in Information Technology with a specialization in Information and Assurance at Capella University. Mr. Hood's technological interests include systems engineering and development, virtualization of system resources, private cloud computing, information assurance and security, and system testing, 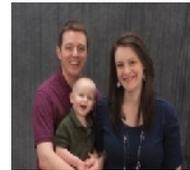 validation, and documentation. Mr. Hood holds a MCP, MCDST, five MCTS, four MCITP, and VCP certifications.

Dr. Syed (Shawon) M. Rahman is an Assistant Professor in the Department of Computer Science and Engineering at the University of Hawaii-Hilo and an adjunct faculty of information Technology, information assurance and security at the Capella University. Dr. Rahman's research interests include software engineering education, data visualization, information assurance and security, web accessibility, and 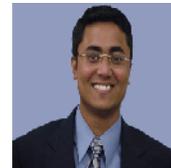 software testing and quality assurance. He has published more than 65 peer-reviewed papers. He is a member of many professional organizations including ACM, ASEE, ASQ, IEEE, and UPE.